\documentclass[12pt]{iopart}
\usepackage{graphicx}

\begin{document}

\title[Instabilities in the Nuclear Energy Density Functional]{Instabilities in the Nuclear Energy Density Functional}

\author{M. Kortelainen and T. Lesinski}
\address{Department of Physics and Astronomy, University of Tennessee,\\
         Knoxville, TN 37996, USA}
\address{Physics Division, Oak Ridge National Laboratory,
         Oak Ridge, TN 37831, USA}
\ead{kortelainene@ornl.gov, tlesinsk@utk.edu}

\begin{abstract}
In the field of Energy Density Functionals (EDF) used in nuclear structure
and dynamics, one of the unsolved issues is the stability of the functional.
Numerical issues aside, some EDFs are unstable with respect to particular
perturbations of the nuclear ground-state density. The aim of this contribution
is to raise questions about the origin and nature of these instabilities, the
techniques used to diagnose and prevent them, and the domain of density
functions in which one should expect a nuclear EDF to be stable.
\end{abstract}

\maketitle

The Energy Density Functional (EDF) method is a broadly used tool in the study
of nuclear structure and dynamics, allowing to describe medium-mass to heavy
nuclei with reduced computational cost. This is achieved thanks to the implicit
resummation of quantum many-body correlations into an effective energy
functional of
local densities, which is, in principle, minimized for the intrinsic density of
the many-body ground state (i.e. the density in the center-of mass frame
\cite{messud}, possibly breaking several symmetries of the Hamiltonian).
Non-relativistic EDFs mostly used in nuclear physics until now are of the Skyrme
type, i.e. semi-local functionals with a postulated analytical form and no clear
connection to the ``bare'' two- and three-nucleon interactions.

The Skyrme EDF has been employed not only in static ``mean-field'' calculations,
but also for the study of small- or large-amplitude collective dynamics in
schemes such as Time-Dependent-Hartree-Fock (TDHF), or Quasiparticle
Random-Phase Approximation (QRPA). Furthermore, the power of the Multi-Reference
EDF method \cite{RMP,duguethere}, which includes collective-motion correlations
into ground and excited states through the resolution of the Hill-Wheeler
equations, has been thoroughly demonstrated.

In the present article we concentrate on ``Skyrme-like'' or semi-local
functionals. One of the intrinsic, unsolved issues of these functionals is the
stability and convergence of calculations performed with them. Besides numerical
issues linked with finding an energy minimum, especially a constrained one, some
functionals appear unstable with respect to specific perturbations of the
nuclear density. This instability means that self-consistent
calculations will yield unphysical configurations at energy minimum, break
symmetries expected in the minimum-energy density (i.e. other than those
supposed to be explicitly broken such as spherical symmetry) and/or yield a
divergent energy. This is now understood to be an intrinsic defect of some
functionals, independently from numerical details \cite{mspli,schunck}.

The Skyrme EDF can be expressed through the local energy density $\mathcal{H}$,
\begin{equation}
\mathcal{H} = \frac{\hbar^{2}}{2m}\tau_{0} +\mathcal{H}_{0}+\mathcal{H}_{1} \, ,
\end{equation}
where isoscalar ($t=0$) and isovector ($t=1$) terms have been separated; both have the same
structure, i.e. for even-even nuclei,
\begin{equation}
\mathcal{H}_{t} = C^{\rho}_{t} \rho_{t}^{2} +C^{\tau}_{t} \rho_{t}\tau_{t}
+C^{\Delta\rho}_{t}\rho_{t}\Delta\rho_{t} +C^{\nabla J}_{t}\rho_{t}\nabla J_{t}
+C^{J}_{t} J^{2}_{t} \,,
\end{equation}
where $\rho$, $\tau$, and $J$ are respectively the matter, kinetic, and
spin-current densities, an index $t=0,1$ indicating respectively isoscalar
and isovector quantities; standard definitions of the latter can be found in
Refs.~\cite{RMP,perlinska}. In typical Skyrme parametrizations, density
dependence is introduced as
$C^{\rho}_{t}=C^{\rho}_{t0}+C^{\rho}_{tD}\rho_{0}^{\gamma}$, with $\rho_0 =
\rho_\mathrm{n} + \rho_\mathrm{p}$. Generalizations
involving a dependence of coupling constants on the isovector density
$\rho_1=\rho_\mathrm{n} - \rho_\mathrm{p}$ and/or more involved functions of
densities are being studied. Recent work \cite{bognerdme,gebremariam} aiming at
building non-empirical, yet semi-local, functionals through the Density Matrix
Expansion (DME) method \cite{negele} also leads to such generalized forms.

The instability of Infinite Nuclear Matter (INM) and nuclei with respect to
global spin and/or isospin polarization is a common problem for functionals
derived strictly as the expectation value of a Skyrme effective interaction, and
has been repeatedly discussed and addressed \cite{blaizot,kutschera,chabanat}.
However, these pathologies of the INM Equation of State (EOS), usually occurring
at high densities, hardly affect nuclear structure calculations in practice.

\begin{figure}
 \begin{center}
    \includegraphics[width=0.7\textwidth]{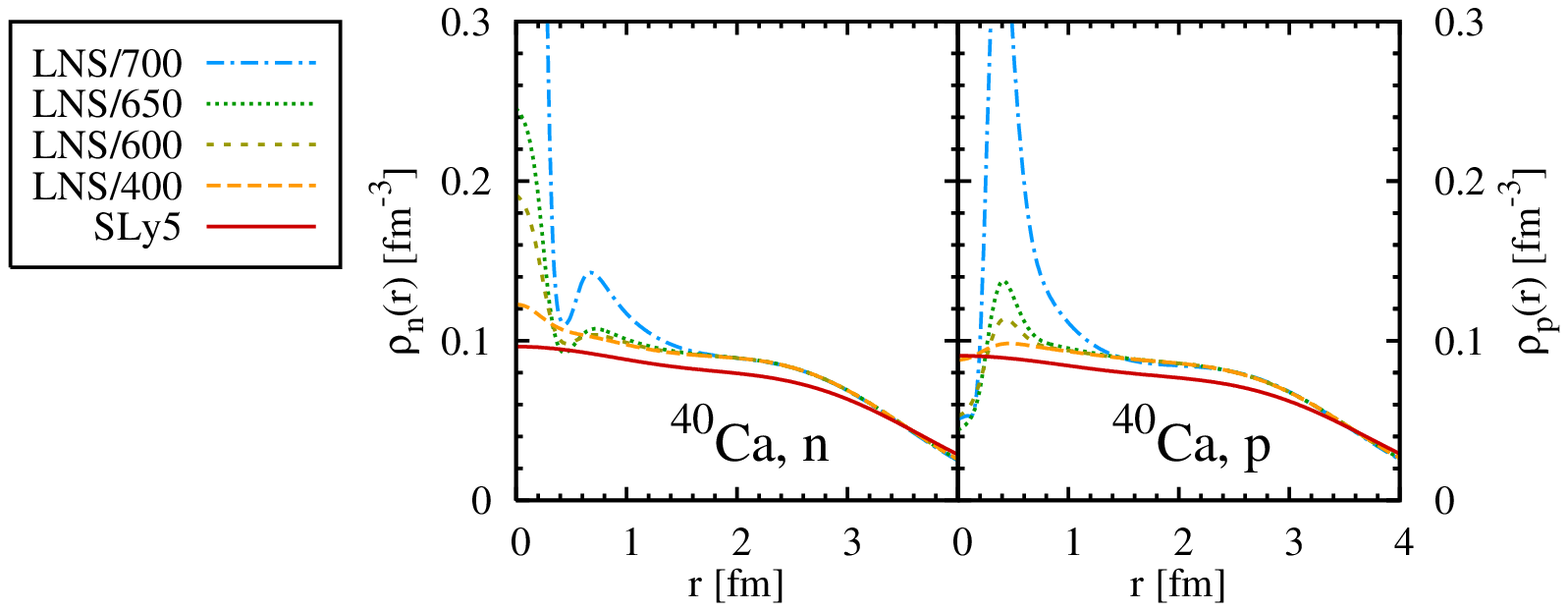}\\
    \includegraphics[width=0.7\textwidth]{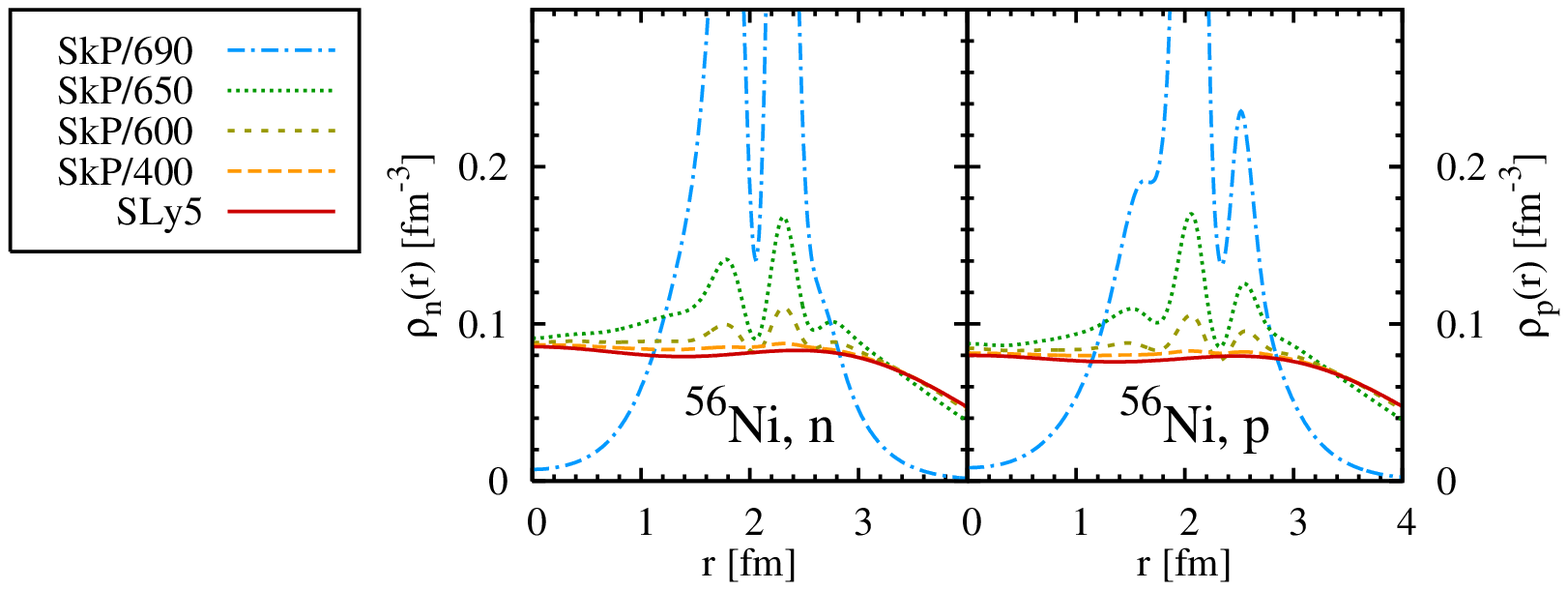}
 \end{center}
 \caption{Radial density profile in nuclei $^{40}$Ca (top panel) and $^{56}$Ni
    (bottom panel) obtained through EDF calculations in spherical symmetry,
    respectively with functionals LNS \cite{cao} and SkP \cite{dft}. Each curve
    corresponds to the number of iterations indicated in the legend. Stable
    results obtained with SLy5 \cite{chabanat2} are given for reference.}
 \label{f:dens}
\end{figure}

More recently, a divergence of the total energy and densities has been observed
in EDF calculations of finite nuclei \cite{mspli} performed with specific
parameterizations of the Skyrme functional \cite{dft,cao}. Fig.~\ref{f:dens}
displays densities obtained with these parameterizations in two sample nuclei
at various (large) numbers of iterations of the self-consistent equations.
Characteristic traits of this phenomenon include unnatural configurations
(large-amplitude spatial oscillations of densities, with separation of protons
and neutrons) and unnaturally large (negative) total energy. This instability
was analyzed by studying the reponse of INM, used as a model system, to
finite-size density perturbations at the Random Phase Approximation (RPA) level.

When calculated with the central terms of the residual interaction deduced from
a semi-local functional \cite{garciarecio}, the analytic expressions for the RPA
response functions are rather involved, even more so when the spin-orbit
\cite{margueron} and tensor \cite{davesne} terms are added. For the sake of
discussion, it is useful to consider the case of a functional with $C^\tau_t =
C^J_t = C^{\nabla J}_t = 0$ (which is strictly local, despite the remaining
$\rho\Delta\rho$, ``finite-range'' terms) considering isospin density
fluctuations only. One then finds the textbook result for the response function
in symmetric INM,
\begin{eqnarray}
 \Pi(\omega,\mathbf{q}) &=& \frac{4\,\Pi_0(\omega,\mathbf{q})}%
    {1-8\,[C^\rho_1 - C^{\Delta\rho}_1 \mathbf{q}^2]\,\Pi_0(\omega,\mathbf{q})}
\end{eqnarray}
where $\omega$ is the excitation energy, $\mathbf{q}$ the transferred
momentum, or wave number of the density fluctuation, $\Pi(\omega,\mathbf{q})$
the response function, and $\Pi_0(\omega,\mathbf{q})$ the non-interacting
response, or Lindhard function \cite{fetter}. The value
$\Pi(\omega=0,\mathbf{q})$ corresponds to the static susceptibility of the
system to finite-size perturbations. It should be positive (with the
above sign convention) for all values of $\mathbf{q}$ and the density $\rho_0$.
(We recall that the coupling constants in the above equation can depend on
the latter, as does $\Pi_0$.) A change of sign with either variable is
accompanied by a pole signaling the existence of a zero-energy collective
mode. We see that the short-wavelength (high $\mathbf{q}$) behaviour is driven
by the coefficient $C^{\Delta\rho}_1$, whose value correlates with the
occurrence rate of the instability observed in calculations of finite nuclei
\cite{mspli}. A similar convergence issue was identified in
Ref.~\cite{zdunczuk}, and confirmed in Ref.~\cite{schunck}, concerning the
susceptibility of the functional SkO \cite{sko} to fluctuations of the spin
density.

We could categorize these instabilities in terms of the importance and role of
the finite size of the system under consideration. First of all, Landau's theory
of Fermi fluids \cite{migdal} gives quantitative constraints for the stability
vs. long-range perturbations of INM in terms of Landau parameters. Second,
the more general random-phase approximation (RPA) calculations in INM can, at
least qualitatively, link zero-energy modes to oscillations of densities in
static calculations of nuclei. Third, in finite nuclei, instabilities
originating at the surface may not be amenable to a diagnosis through INM
calculations. This may be emphasised, in particular, by the recent attention
brought to more involved density dependences \cite{mizuyama}. This leads to ask
a first open question:

\begin{enumerate}
\item[1] Can we quantitatively understand the link between instabilities in INM 
and finite nuclei? In particular, what INM densities are relevant to the study
of nuclei? How does finite size modify the results from INM?
\end{enumerate}

\begin{figure}[tbh]
 \begin{center}
  \includegraphics[width=5cm]{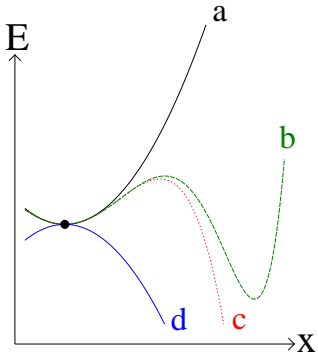}
 \end{center}
 \caption{Schematic picture of energy landscape with respect to
    collective coordinate $x$, assumed to be constrained while other degrees 
    of freedom are variationally optimized, for stable (a) and various unstable
    cases. See text.}
 \label{f:scheme}
\end{figure}

The principal manifestation of instabilities described up to now seems to be
divergence of given quantities. However, let us point out that this is not
mandatory for a functional to exhibit unstable behaviour. A schematic
representation of the EDF energy with respect to an abstract collective
coordinate $x$ in the EDF calculation (e.g. in Fig.~\ref{f:dens}, the magnitude
of the density oscillations) is given in Fig. \ref{f:scheme}. Here the dot at
$x=0$ represents the physically expected configuration. Curve (a) corresponds to
the (desired) situation with a stable configuration, whereas curve (d)
represents a well-developed instability. Curves (b) and (c) represent situations
with a metastable minimum which can be made to evolve to more energetically
favored configurations by some perturbation. In case (b) a true, global
variational solution exists; whereas in case (c) it does not. Here, RPA
(i.e. linear response) analysis would only warn about situation (d).

Case (b) was encountered in Appendix~B of Ref.~\cite{tensor}, where a
parameterization with large negative values of $C^J_t$ was found to yield, in
seemingly random nuclei, an unphysical shell structure. The plot of the energy
landscape as a function of the magnitude of the spin-current density $J$ in
Fig.~35 of Ref.~\cite{tensor} indeed corresponds quite well to case (b),
with two minima, the unphysical one being energetically favored. Thus, it seems
appropriate to ask:

\begin{enumerate}
\item[2] How can we diagnose instabilities not accessible with a small-amplitude
expansion around the expected ground state (as in RPA)?
\end{enumerate}

In Ref.~\cite{maruhn}, a TDHF calculation of a central $^{16}$O-$^{16}$O
reaction done with the functional SkM$^\ast$~\cite{bartel} showed that after
contact, the two nuclei separated exhibiting a toroidal spin polarization. This
puzzling result might be analyzed from the standpoint of stability analysis: is
this configuration lying lower in energy than a more ``reasonable'' one? If
so, this result might well have to be discarded. It is also
possible that a treatment beyond the static considerations above (i.e. referring
to energy minimization) be necessary for a thorough understanding, and hence:

\begin{enumerate}
\item[3] What are the manifestations of instabilities developing on top of
excited states, probed in dynamical calculations? Can we reduce their study to
a model problem, such as the INM analysis above?
\end{enumerate}

Finally, current efforts tend towards producing non-empirical semi-local
functionals through the DME applied to low-momentum interactions \cite{bogner}.
The DME, more than a simple range expansion, consists of a series of finite-size
corrections to the INM density matrix \cite{negele,bognerdme,gebremariam}. It
can thus be expected to be accurate for smoothly varying densities such as those
occurring naturally in nuclei. However, there is no guarantee that this accuracy
will be preserved for pathologic configurations such as in Fig.~\ref{f:dens}. In
other words, extra care should be taken to verify that the DME functional does
not exhibit pathological behaviour beyond its expected domain of validity.
Indeed, by going from an exact Hartree-Fock functional deduced from a realistic
Hamiltonian to a semi-local functional, spurious high-momentum components are
introduced e.g. in the residual interaction and response functions (Eq.~(3)).
The apparent link between RPA linear response analysis and self-consistent EDF
calculations suggests that the ``singularity'' due to the semi-local nature of
the functional can make divergences, naturally expected in calculations beyond
first order of perturbation theory performed with a zero-range interaction,
creep into ``mean-field'' ones through self-consistency. Our final question is
thus:

\begin{enumerate}
\item[4] How far in density should we expect the next-generation nuclear EDF
to be stable, and in what domain of variation for the reference Slater
determinant and associated densities? In other words, shouldn't semi-local
functionals be specified with a well-defined ``cutoff'' or resolution scale?
\end{enumerate}

It is especially important to understand this issue for functionals containing
higher-order derivative terms (yielding higher orders of $\mathbf{q}$ in
Eq.~(3)) such as proposed in Ref.~\cite{carlsson}, whether they are built from
an underlying Hamiltonian or purely phenomenological.

In summary, despite their great usefulness, semi-local nuclear energy density
functionals of the Skyrme type come with several kinds of stability-related
issues, which can compromise the significance of results obtained with them.
Indeed, whereas in the process of developing new nuclear functionals, one
naturally focuses on minimum (ground-state) energies and associated density
configurations, these functionals will be probed in different ways in actual
applications, be it during energy minimization or in dynamical extensions of the
theory. These issues should thus be fully understood and kept in mind for future
development.

\section*{Acknowledgements}

We acknowledge fruiful discussions with M. Stoitsov and N. Schunck.
This work was supported by the U.S. Department of Energy
under Contract Nos. DE-FG02-96ER40963, DE-FC02-07ER41457, DE-FC02-09ER41583
(UNEDF SciDAC Collaboration), and DE-FG02-07ER41529
(University  of Tennessee).

\section*{References}

\bibliographystyle{unsrt}
\bibliography{openst-instab}

\end{document}